\documentstyle[multicol,eqsecnum,aps,prd,epsf]{revtex}
\def\dA{d_{\rm A}}
\def\rH{r_{\rm H}}
\def\ms{m_{\rm s}}
\def\mg{m_{{\rm g} i}}
\def\bM{\bar M}

\def\half{{1\over 2}}
\def\NH{N_\protect{\rm H\protect}}
\def\NT{N_\protect{\rm T\protect}}

\def\Hsi{H_\protect{\rm shell\protect}}
\begin{document}
 \makeatletter
 \newdimen\ex@
 \ex@.2326ex
 \def\dddot#1{{\mathop{#1}\limits^{\vbox to-1.4\ex@{\kern-\tw@\ex@
  \hbox{\tenrm...}\vss}}}}
 \makeatother
\thispagestyle{empty}
{\baselineskip0pt
\leftline{\large\baselineskip16pt\sl\vbox to0pt{\hbox{Department of Physics} 
               \hbox{Kyoto University}\vss}}
\rightline{\large\baselineskip16pt\rm\vbox to20pt{\hbox{KUNS-1550} 
\vss}}%
}
\vskip1cm
\begin{center}
{\large{\bf Distance-redshift relation in an
isotropic inhomogeneous universe II}:}\\
Spherically symmetric dust-shell universe \\
\end{center}
\begin{center}
{\large Norimasa Sugiura\footnote{Email address:
    sugiura@tap.scphys.kyoto-u.ac.jp}, 
Ken-ichi Nakao\footnote{Email address: nakao@tap.scphys.kyoto-u.ac.jp}
and Tomohiro Harada\footnote{Email address:
  harada@tap.scphys.kyoto-u.ac.jp}}\\ 
{\em Department of Physics,~Kyoto University, 
 Sakyo-ku, Kyoto 606-8502, Japan}
\end{center}
\begin{abstract}
The relation between the angular diameter distance and redshift 
($\dA$-$z$ relation) 
in a spherically symmetric dust-shell universe is studied.
We have discovered that the relation agrees with that 
of an appropriate Friedmann-Lema\^{\i}tre (FL) model 
if we set a ``homogeneous'' expansion law and a ``homogeneous''
averaged density field. 
This will support the averaging hypothesis that a universe
looks like a FL model in spite of small-scale fluctuations
of density field,
if its averaged density field is homogeneous on large scales.
We also study the connection of the proper mass of a shell
with the mass of gravitationally bound objects.
Combining this with the results of the $\dA$-$z$ relation,
we discuss an impact of the local inhomogeneities
on determination of the cosmological parameters
through the observation of the locally inhomogeneous universe.
\end{abstract}


\section{Introduction}
The standard big bang model is based on the assumption of 
the homogeneous and isotropic distribution of matter and 
radiation. This assumption then leads to the Robertson-Walker 
(RW) spacetime geometry and the Friedmann-Lema\^{\i}tre (FL) 
universe model\footnote{We use the term ``Robertson-Walker
spacetime'' when we focus on geometrical aspects of a homogeneous
and isotropic model, and ``Friedmann-Lema\^{\i}tre model'' when we
discuss its dynamics and observable quantities.}
through the Einstein equation. 
This standard model has succeeded in explaining various important
observational facts: Hubble's expansion law, the content of light
elements and the existence of the cosmic microwave background
radiation (CMBR)\cite{rf:WEINBERG}. 

The CMBR conversely gives a strong 
observational basis for the assumption of homogeneity and 
isotropy of our universe by its highly isotropic 
distribution together with 
the assumption that we are not in any special position 
in the universe (the Copernican principle). 
In fact, the deviation 
of our universe from a homogeneous and isotropic space 
is as small as $\sim 10^{-5}$
at the stage of decoupling\cite{rf:COBE}.
Thus our universe is well approximated by 
a FL model before this stage. 
On the other hand, the present universe is highly inhomogeneous 
on small scales; the density contrast against the 
cosmic background density is of the order of $10^{30}$ for the sun,
$10^{5}$ on galactic scales,
and of the order of unity even on the scale of superclusters. 
We have to go beyond FL models and linear perturbations
in considering such systems.

We usually regard that
a FL model is a large-scale ``average'' of a
locally inhomogeneous universe (averaging hypothesis). 
Even though the observational data are consistent with the 
picture that our universe is described well by 
a RW metric with small perturbations, we are not sure 
how to derive the background FL model from the inhomogeneous universe
by any averaging procedure,
or how the non-linear inhomogeneities on small scales 
affect large-scale behavior of the universe\cite{rf:ELLIS}.
Although one can derive a background 
FL model from observations of the nearby galaxies
with any rule of averaging one likes,
it is uncertain
whether this background FL model (or its time evolution backward)
agrees well with the highly homogeneous universe
at the early regime.
The discrepancies might appear, for example, in the 
estimate of the density of baryonic matter, 
the density parameter, the age of our universe, and so on.
These still remain a non-trivial question 
to which we have to give a clear answer.

Before proceeding to discuss these problems,
we should make the meaning of the term ``average'' definite.
Averaging can be defined as a mapping
between an inhomogeneous spacetime and a homogeneous one.
This mapping is not restricted to simple volume averaging;
other averaging methods such as deformation of 
three-geometry according to the Ricci-Hamilton flow may be
possible\cite{rf:CP}. 
By averaging, anyway,
we expect that the large-scale 
(or coarse grained) behavior of the inhomogeneous spacetime
is extracted.

Averaging problem has been often studied from the
viewpoint of the so-called back-reaction problem, i.e.,
how the small-scale  
inhomogeneities affect the global dynamics when averaged 
on larger scales.
Using the perturbation formalism and volume averaging,
the back-reaction problem has been studied by several authors
\cite{rf:FUTAMASE,rf:BE,rf:RSKB,rf:BOERSMA,rf:TANIMOTO}.
Apart from the problem whether the volume averaging is 
appropriate or not,
one reason which makes the discussion of averaging unclear is
that there exists no natural choice of time slice
in an inhomogeneous universe.
This leads to an ambiguity in the definition of averaged 
expansion rate or averaged density.
Actually, there are even apparent discrepancies in the statements
of the papers above.

One possible prescription to avoid this difficulty
is to treat observable quantities 
which we can give a clear definition. From 
this viewpoint, observational effects on the CMBR of 
small-scale inhomogeneities have been studied 
\cite{rf:BILD,rf:BF}. 
These are significant in the sense that the back reactions 
on the observable quantities were discussed
(see, e.g., \cite{rf:MBHE} for a recent discussion of observable
quantities in the Lema\^{\i}tre-Tolman-Bondi spacetime).
However, 
the problem how to determine the cosmological parameters 
by observing an inhomogeneous universe
seems to have been overlooked.

In Sugiura {\it et al.}(1998; hereafter Paper I)\cite{rf:PAPER1}, 
we investigated the  
distance-redshift relation in a spherically symmetric
dust-shell model, and compared it with that of a FL model.
We focused on a highly (locally) inhomogeneous model;
we prepared pressureless fluid distributed in discrete shells,
which cannot be described by small perturbations of a FL background.
We discussed the relation between the behavior of the $\dA$-$z$
relation and the conditions on the initial time slice
concerning the mass density and the expansion rate. 
We found that 
the distance-redshift relation observed at the center
obeys a FL-like relation, 
even when there exist only several shells in the initial
horizon scale,
if the following conditions are satisfied on the initial spacelike
hypersurface: 
the expansion law is homogeneous and 
the density which is averaged on larger scales than the inhomogeneity scale
is scale-independent
(i.e., large-scale homogeneity of density field).
Here the averaged density is defined simply by dividing 
the mass of shells by the volume of the hypersurface. 

Our previous analysis was, however, limited to a spatially flat case
(the Einstein de-Sitter model). 
There is a claim that the Einstein-de Sitter solution
is in a special position as a solution of the Einstein equation;
it is a kind of fixed point under renormalization 
group flow defined by the scaling properties of the Einstein equation
\cite{rf:SOTA}.
Thus, there is a possibility that our results were due to the 
special behavior of the perturbations of the Einstein-de Sitter model
under an averaging procedure. 
In this paper, we treat non-flat cases
and show that statements similar to those of our previous paper
do hold.
We also answer some questions unsolved in Paper I.
In particular, 
the previous study of spatially flat cases could not answer
which mass should be used in defining the average
density, proper mass which appears in the stress-energy
tensor of dust shell,
or gravitational mass which specifies the parameter of the 
Schwarzschild spacetime; they are the same in the spatially flat model.
Also it will be shown that the significance of 
the effect of inhomogeneities has curvature dependence.

We here note three interesting points of the dust-shell model.
First, its dynamics is exactly solved;
it is not necessary to assume the existence of 
homogeneous background in order to obtain the evolution
of matter distribution.
Secondly, it treats a discrete mass distribution 
where the linear perturbation theory is invalid, and 
also can treat a highly general-relativistic situation 
where the scale of inhomogeneities are comparable to 
the horizon scale. 
In a sense, sphericall dust-shell model is 
a limit case of the Tolman-Bondi solution.
We expect it can represent the Tolman-Bondi and the FL solutions 
when we take appropriate limits where the number density of dust shell 
goes to infinity, though no rigorous proof has been yet obtained.
Thirdly,
the motion of each shell can be extended even after shell-crossing occurs.
This theme is discussed elsewhere\cite{rf:IDA,rf:NAKAO}.

In order to get an insight into the inhomogeneous universe 
from the study of the dust-shell universe, 
we will also investigate the origin of the dust shell. 
Assuming that the dust shell is composed of highly 
bound objects, we consider the effect of the binding energy of each 
object on the dynamics of the universe.

The organization of this paper is as follows. 
In the next section, we summarize the basic equations 
for the dynamics of a dust-shell universe,
distance to the shells from the center,
and redshift of the shells measured by an observer at the center.
This is an extension of the treatment in Paper I.
We give our results and discussion on $\dA$--$z$ relation 
and averaged density in Sec. III. 
In this section, we also discuss the universe filled with 
gravitationally bound objects
and the effect of the binding energy of each object on the dynamics of the 
universe. Finally Sec. IV is devoted to the summary.

We follow the sign convention of the Riemann tensor and
the metric tensor in \cite{rf:WALD} 
and adopt the unit of $G=c=1$.

\section{Formulation of Dust-shell universe}
\subsection{Equation of motion of dust shell}
First, we derive the expansion law of the dust-shell universe.
We consider a number of spherically symmetric shells
with a common center at $r=0$ (see Fig. 1). 
We label each shell by $1,2,\cdots$,$i,\cdots$ from inside.
The region enclosed by the first shell is the Minkowski spacetime
and is labeled as the first region.
Similarly, the region enclosed by the $(i-1)$th shell and $i$th shell 
, which is a vacuum, is labeled as the $i$th region. 
Each shell is infinitesimally thin, characterized by the 
surface stress-energy tensor $S^{ab}$ which is given by
\begin{eqnarray}
  S^{ab} \equiv \lim_{\epsilon\rightarrow 0} 
\int_{-\epsilon}^{+\epsilon}
 T^{ab}\ dx ,
\end{eqnarray}
where $x$ is a Gaussian coordinate ($x=0$ on the shell) in 
the direction normal to the trajectory of the shell.

Since each region between the shells is a vacuum, 
the spacetime is described by the Schwarzschild geometry 
by Birkhoff's theorem. The line element in the $i$th 
region is written in the form
\begin{eqnarray}
  ds_{i}^2 = - \left( 1 - {2m_{{\rm g}i}\over r}\right) dt^2 
  + \left( 1 -{2m_{{\rm g}i}\over r}\right)^{-1} dr^2 
  + r^2 d\Omega^2, \label{eq:st-line}
\end{eqnarray}
where the parameter $m_{{\rm g} i}$ will be 
referred to as a gravitational mass ($m_{{\rm g}1} = 0$), and
$d\Omega^2$ is the line element of a unit sphere.

We derive 
the expansion law of a dust shell
following\cite{rf:MTW,rf:ISRAEL,rf:SATO,rf:MAEDA}.
Let $n^a$ be a unit spacelike vector normal to the trajectory of the
shell, and 
define the projection operator $h_a^b \equiv \delta_a^b - n_a n^b$. 
From the projected Einstein equation 
\begin{eqnarray}
R_{ab} h^a_c h^b_d  =  8\pi  \left( T_{ab} - {1\over 2}g_{ab} T\right)
h^a_c h^b_d ,
\end{eqnarray}
one obtains 
\begin{eqnarray}\label{proj}
  {\pounds}_n K_{cd} + {}^3 R_{cd} - K K_{cd} =
   8\pi  \left(T_{ab}h^a_c h^b_d - {1\over 2}h_{cd} T \right) ,
\end{eqnarray}
where ${\pounds}_n$ is the Lie derivative along $n^a$
and ${}^3 R_{cd}$ is the three-dimensional Ricci tensor of the 
timelike hypersurface generated by the motion of the shell.
The extrinsic curvature $K_{ab}$ is defined by 
$K_{ab} = - {1\over 2} h_a^c h_b^d
{\pounds}_n h_{cd}$, and $K = K_a^a$, $T= T_a^a$.
Integration of Eq. (\ref{proj}) 
over an infinitesimal range along $n^a$ yields
\begin{eqnarray}\label{K1}
  K_{ab}^+ - K_{ab}^- =  8\pi 
\left( S_{ab} - {1\over 2}h_{ab} S\right) ,
\end{eqnarray}
where the suffix `$+$' denotes a quantity evaluated at the outside of
the shell, and `$-$' at the inside.
Using Eq. (\ref{K1}) and the Gauss-Codazzi relation
$  2G_{ab} n^a n^b = - {}^3 R + K_{ab} K^{ab}- K^2 $,
one finds that the following relation holds for a dust shell:
\begin{eqnarray}\label{K2}
  S^{ab} \left( K_{ab}^+ + K_{ab}^- \right) = 0 .
\end{eqnarray}
The stress-energy tensor of a dust shell is given by 
\begin{eqnarray}\label{s-e}
  S^{ab} = s u^a u^b,
\end{eqnarray}
where $u^a$ is the 4-velocity of the observer 
rest on the shell, and 
$s$ is the surface energy density of the shell.
Combining Eqs. (\ref{K1}), (\ref{K2}) and (\ref{s-e}),
we obtain the following equation for the circumferential radius $r_i$
of the $i$th shell (the ``expansion law'' of the dust shell):
\begin{eqnarray}\label{motion}
 \left({dr_{i}\over d\tau}\right)^2 = {2\bM_i\over r_i} 
+ \left\{ \left({M_-(i)\over m_{\rm s}(i)}\right)^2 -1\right\}
+ {  m_{\rm s}^2(i)\over 4 r_i^2} ,
\end{eqnarray}
where $\bM_i$ and $M_{-}(i)$ are defined by
\begin{eqnarray}
  \bM_i \equiv {\mg + m_{{\rm g} i+1}\over 2} , \label{eq:gmass-def}
\end{eqnarray}
\begin{eqnarray}
  M_- (i) \equiv  m_{{\rm g}i+1}-\mg  ,\label{eq:dmass-def}
\end{eqnarray}
and $\tau$ is the proper time which is 
measured by an observer rest on the shell.
The ``proper mass'' $\ms (i)$ of the shell is defined by 
\begin{eqnarray}\label{eq:rmass-def}
m_{\rm s}(i) = 4 \pi s_i r_i^2 
\end{eqnarray}
where $s_i \equiv - S^a_a (i)$.
It can be shown this proper mass is a constant of motion 
by the conservation law, ${S_a^b}_{;b} = 0$, where 
the semicolon denotes the three-dimensional covariant derivative on the 
trajectory of the shell. 
We shall assume the energy condition $m_{\rm s} \geq 0$.

In this paper, we use a common proper time $\tau$ for 
all the shells. The relation between $\tau$ and 
the time coordinate, $t$, in the Schwarzschild spacetime is obtained as
follows. First, note that two Schwarzschild time coordinates
are assigned to  
each shell: the $i$th shell has the time $t_{(-)i}$ measured 
in the $i$th region and $t_{(+)i}$ measured in the $(i+1)$th 
region. 
From Eqs.(\ref{K1}) and (\ref{K2}), we obtain
\begin{eqnarray}
 {dt_{(+)i}\over d\tau}
&=&\left({r_{i}\over r_{i}-2m_{{\rm g}i+1}}\right)
  \left[ {M_-(i)\over m_{\rm s}(i)}-{m_{\rm s}(i) \over 2r_i}
  \right], \label{eq:ex-time}\\
 {dt_{(-)i}\over d\tau}
&=&\left({r_{i}\over r_{i}-2\mg}\right)
  \left[
 {M_-(i)\over m_{\rm s}(i)}+{m_{\rm s}(i) \over 2r_i}
  \right]\label{eq:in-time}.
\end{eqnarray}
The procedure to determine the origin of each time coordinate will be 
described later. 

\subsection{Cosmological parameters and initial condition}
In order to specify a dust-shell universe, we have to fix
the parameters in Eq. (\ref{motion}) and the initial hypersurface.
We first rewrite Eq. (\ref{motion}) in the form
corresponding to the Hubble equation of a FL model.
We denote the initial circumferential radius of the $i$th shell 
by $x_{i}$, i.e., 
\begin{equation}
r_{i}=x_{i}
\end{equation}
on initial hypersurface.
We define $\rho_i$ by
\begin{eqnarray}
 \bM_i \equiv {4\over 3}\pi \rho_i {x_i}^3 ,
\label{eq:g-mass}
\end{eqnarray}
and $k_{i}$ by
\begin{eqnarray}\label{eq:curvature}
{M_-(i)\over m_{\rm s}(i)}
\equiv {E_i}
\equiv \sqrt{1-k_{i}x_{i}} .
\end{eqnarray}
Here ${E_i}$ is the specific energy of the $i$th shell.
$E_{i}$ is positive through the space in the open and flat FL models
and inside the maximum radius in the closed model.
In this paper, we consider the cases where $E_i > 0$.

Using these parameters, the expansion law of the dust shell 
can be written as 
\begin{eqnarray}\label{expa}
 \left( {1\over r_{i}}{dr_{i}\over d\tau}\right)^2 
 = {8\over 3} \pi  \rho_i
 \left( {x_i \over r_i}\right)^3  
 -k_{i}\left({x_{i}\over r_i}\right)^{2}
 + {  m_{\rm s}^2(i)\over 4 r_i^4}  .
\end{eqnarray}
We see that the first term behaves like a non-relativistic matter term
in the Hubble equation of a FL model, the second and the third
like a curvature and a radiation source term.
From this point of view, $\rho_i$ and $k_i$ play roles
of the ``energy density'' and the ``curvature'', respectively.
The radiation-like term might be regarded as the effect 
of the binding energy of the shell\cite{rf:SATO}. 
Further it is worthwhile to note that this radiation-like term 
is consistent with the result of Futamase\cite{rf:FUTAMASE}
about the effect of the small-scale
inhomogeneities on the global cosmic expansion.
Seeing this, one may expect that the inhomogeneities tend to make 
the Hubble parameter larger 
compared with a homogeneous universe
which has the same ``energy density'' of non-relativistic matter.
However, this radiation-like term 
does not necessarily imply the larger Hubble parameter.
In order to see the effect of this term
on the Hubble parameter, we need to investigate 
the distance-redshift relation 
by solving the null geodesic equations and compare 
the result in the inhomogeneous spacetime and that 
of the FL model. We will discuss this point later.

A dust-shell universe is specified
if we set the parameters contained in Eq. (\ref{expa}), i.e.,
$\rho_i$, $k_i$, $x_i$, and an initial hypersurface.
When we increase the number of the shells to infinity
with $\rho_i$ and $k_i$ being finite and independent of $i$
(we will mention this limit as ``large $N$ limit''),
the dust-shell universe approaches a FL universe
if we take an appropriate initial hypersurface, as we will see in
Sec. III.
Then the parameters $\rho_i$ and $k_i$ agree with the ordinary
energy density and curvature in the Hubble equation.

We need to derive the expression of $t_{(\pm)i}$ in terms of $r_{i}$ 
for the later use.
The relations between 
$t_{(\pm)i}$ and $r_{i}$ are given by [we denote $m_{\rm s}(i)$ as
$m_i$ in the remainder of this section] 
\begin{equation}
  {dt_{(\pm)i}\over d r_i}
  = {\xi_{i}( E_i {r_i}^2 \mp {1\over 2} m_i r_i )\over
    ( r_i- 2{\bar M}_i \mp E_i m_i) \sqrt{X_i (r_i)}},
  \label{eq:t-r-eq}
\end{equation}
where
\begin{eqnarray}
  X_i (r_i)=({E_i}^2 -1) {r_i}^2 + 2{\bar M}_i r_i +{{m_i}^2\over 4},
\end{eqnarray}
and $\xi_{i}$ is the sign of $d r_i/d\tau$.
This is integrated to give
\begin{equation}
  t_{i}|_{\pm} = \xi_i T_{i\pm}(r_i) +\xi_i\left(2{\bM}_i\pm E_i m_i\right) 
    \ln\biggl|{2{\bM}_i\over G_{i\pm}(r_i)}
    \left(r_i-2{\bar M}_{i}\mp E_i m_i\right)\biggr|
  +{\cal T}_{i\pm},\label{eq:time-sol}
\end{equation}
where ${\cal T}_{i\pm}$ are integration constants.
The functions  $T_{i\pm}(r)$ and $G_{i\pm}(r)$ 
are given for $E_{i}^{2}=1$ as
\begin{eqnarray}
 T_{i\pm}(r)
 &=&{1\over 3{\bM_i}^{2}}
    \left({\bM_i} r +6{\bM_i}^{2}\pm {3\over 2}{\bM_i} m_i 
      - {m^{2}_i\over 4}\right)
    \sqrt{X_i (r)}, \\
 G_{i\pm}(r)
 &=&\left\{\sqrt{X_i(r)}+2{\bM_i}\pm {m_i\over 2}\right\}^{2}, 
\end{eqnarray}
for $E^{2}>1$ as
\begin{eqnarray}
 T_{i\pm}(r)
 &=&{E_i \over {E_i}^2-1} \sqrt{X_i(r)} 
    +\left\{ {  E_i (2{E_i}^2 -3) {\bM_i}
               \pm(2{E_i}^2 -1)({E_i}^2 -1){m_i \over 2}
     \over (E_i^2 -1) \sqrt{E_i^2-1}}
     + 2 {\bM_i} \pm E_i m_i \right\} \\ \nonumber
 &{}&\times  \ln\left[ {\bM}_i + ({E_i}^2 -1)r + \sqrt{{E_i}^2 -1}
                \sqrt{X_i(r)}
                \right] ,\\
 G_{i\pm}(r) &=& 
 \left\{ \sqrt{{E_i}^2 -1}\ r + \sqrt{X_i(r)} + 
         2\left(E_i-\sqrt{{E_i}^2 -1}\right) {\bM}_i
         \pm{1\over 2}\left(E_i-\sqrt{{E_i}^2 -1}\right)^2 m_i
 \right\}^2,
\end{eqnarray}
and for $E^2 < 1$ as
\begin{eqnarray}
 T_{i\pm}(r)
  &=& {E_i \over {E_i}^2-1} \sqrt{X_i(r)}
    +\left\{ {  E_i (2{E_i}^2 -3) {\bM_i}
               \pm(2{E_i}^2 -1)({E_i}^2 -1){m_i \over 2}
     \over (E_i^2 -1) \sqrt{1-E_i^2}}\right\}
     {2} {\rm ArcTan}\left( \sqrt{r+k\over h-r}\right) ,\\
 G_{i\pm}(r) 
  &=& \left\{ \sqrt{(h-2{\bM}_i \mp E_i m_i)(k+r)} 
               + \sqrt{(k+2{\bM}_i\pm E_i m_i)(h-r)}
      \right\}^2,
\end{eqnarray}
where $h$ and $-k$ are the roots of $X_i(r) =0$;
\begin{eqnarray}
  h 
  &=&{{\bar M}_i+\sqrt{ {{\bar M}_i}^2+(1-E_i^2)m_i^2 /4} \over (1-E_i^2)} ,\\
  k 
  &=&-{{\bar M}_i-\sqrt{{{\bar M}_i}^2+(1-E_i^2)m_i^2 /4} \over (1-E_i^2)} .
\end{eqnarray}
There is a coordinate singularity on the Killing horizon
in the Schwarzschild coordinate; 
$t_{(\pm)i}$ becomes infinite on $r_{i}=2 \mg$. 
In order to avoid this coordinate singularity, and
for further calculation, the Kruskal null coordinate is convenient 
since we will treat the null geodesics 
in this spacetime.
Outside the horizon in the $i$th region, $r>2\mg$, 
the Kruskal null coordinates are given by
\begin{eqnarray}
  U &\equiv& -4\sqrt{ \mg (r - 2 \mg)}
                   \exp\left({r-t \over 4\mg} \right), \\
  V &\equiv& +4\sqrt{ \mg (r - 2 \mg)}
                   \exp\left({r+t \over 4\mg} \right),
\end{eqnarray}
where $U$ and $V$ correspond to the retarded time and
 the advanced time, respectively. Using these Kruskal coordinates, 
the line element in the $i$th region is expressed as
\begin{equation}
  ds_{i}^{2} = -{\mg \over r} \exp\left( -{r \over 2\mg} \right)dUdV
               +r^{2}d\Omega^{2}.
\end{equation}
Similarly to the Schwarzschild time coordinate,
two pairs of Kruskal null coordinates, 
$U_{(\pm)i}$ and $V_{(\pm)i}$, are assigned to each shell.
Using Eq. (\ref{eq:time-sol}), 
we obtain the Kruskal null coordinates labeling the 
$i$th shell in the form
\begin{eqnarray}
 U_{(\pm)i}(r_{i})
 &=& -4 \sqrt{\bM_i \pm {E_i\over 2} m_i} ({2\bM_i})^{-\xi_i/2}
       (r_i - 2\bM_i \mp E_i m_i)^{\half(1-\xi_i)}
       {G_{i\pm}}^{\half \xi_i}(r_i)
       \exp\left[ {r_i - T_{i\pm}(r_i) - {\cal T}_{i\pm} \over
                  4\bM_i\pm 2 E_i m_i} \right] , \label{def:U}\\
 V_{(\pm)i}(r_{i})
 &=& +4\sqrt{ \bM_i \pm {E_i\over 2} m_i }({2\bM_i})^{\xi_i/2}
       (r_i - 2\bM_i \mp E_i m_i)^{\half(1+\xi_i)}
       {G_{i\pm}}^{-\half\xi_i}(r_i)
       \exp\left[ {r_i + T_{i\pm}(r_i) + {\cal T}_{i\pm} \over
                  4\bM_i\pm 2 E_i m_i} \right] .\label{def:V}
\end{eqnarray}
These coordinates, $U_{(\pm)i}$ and $V_{(\pm)i}$, are finite 
on $r_{i}=2\mg$ and are well defined 
also for $r_{i}<2\mg$.  
Hence, we will use the expressions (\ref{def:U}) and (\ref{def:V})
for any $r_i$.
For $\xi_i =+1$, 
both $U_{(\pm)i}$ and $V_{(\pm)i}$ are negative
when $2\mg$ is larger than $r_{i}$.
In this case, the $i$th shell with $r_{i}<2\mg$ 
is located in the white hole part of the Schwarzschild 
spacetime. 
On the contrary, for $\xi_i =-1$, the $i$th shell with $r_{i}<2\mg$ 
is located in the black hole part.

The determination of the integration constants ${\cal T}_{i\pm}$
corresponds to the choice of the initial hypersurface. 
The procedure to construct the initial hypersurface we adopt
is summarized as follows.
First, we choose a unit spacelike vector
$\ell^a$  
which is directed outward (i.e., pointing 
 the direction in which label $i$ increases).
Taking this vector as a starting tangential vector,
we extend a spacelike geodesic curve until it reaches 
the second shell.
This spacelike geodesic curve defines the simultaneous hypersurface 
in the region between the first shell and the second shell.
Next we extend from this intersection towards the third shell
another spacelike geodesic 
which starts with another spacelike vector at the second shell.
This second spacelike geodesic 
generates a spacelike hypersurface in this region. 
Repeating this process, we complete the whole initial hypersurface.

From the above procedure, the integration constant of Eq.
(\ref{eq:time-sol}) is 
determined as follows. In 
the $i$th region,  
the unit tangent vector, $\ell^a$, of the spacelike geodesic is written as
\begin{equation}
  \ell^{t}=e_{i}\left(1-{2\mg\over r}\right)^{-1}, ~~~
  \ell^{r}=\sqrt{1+e_{i}^{2}-{2\mg\over r}}, \label{eq:sg-comp}
\end{equation}
and the other components vanish, where $e_{i}$ is an integration 
constant associated with the geodesic equation.
Now a choice of initial time slice reduces to a choice of $e_i$.
In Sec. III, we will see the connection between the choice of $e_i$
and the $\dA$-$z$ relation.
Once we fix $e_i$, the equation for the 
trajectory of the spacelike geodesic in the $(t,r)$ plane 
is given by
\begin{equation}
  {dt\over dr}={e_{i}r^{3\over2} \over 
                 (r-2\mg)\sqrt{(1+e_{i}^{2})r-2\mg}}  .
\end{equation}
Integrating the above equation, we obtain
\begin{equation}
  t=F_{i}(r)+D_{i},
\end{equation}
where 
\begin{eqnarray}
  F_{i}(r)=
        {e_{i}\sqrt{r}\sqrt{(1+e_{i}^2)r-{2\mg}}\over 1+e_{i}^2}
       + {2\mg} \ln\left({
           \sqrt{(1+e_{i}^2)r-{2\mg}}-e_{i}\sqrt{r}\over
             \sqrt{(1+e_{i}^2)r-{2\mg}}+e_{i}\sqrt{r}}\right)
        \nonumber \\
       +
        {{2\mg} e_{i}(3+2e_{i}^2)
          \ln(\sqrt{(1+e_{i}^2)r-{2\mg}}+\sqrt{(1+e_{i}^2)r})\over 
            (1+e_{i}^2)^{3/2}}   ,
\end{eqnarray}
and $D_{i}$ is an integration constant. 
Initially, we set $t_{(+)i}=t_{(-)i}$ and 
$t_{(\pm)1}=0$. 
Using these relations, we obtain the integration constants, 
${\cal T}_{(\pm)i}$, 
as
\begin{equation}
  {\cal T}_{(\pm)i} = \sum_{j=2}^{i}
        \left[F_{j}(x_{j})-F_{j}(x_{j-1})\right] - \xi_i T_{i\pm}(x_i) 
         -\xi_i\left(2{\bM}_i\pm E_i m_i\right) 
         \ln\biggl|{{\bM}_i\over G_{i\pm}(x_i)}
         \left(x_i-2{\bM}_{i}\mp E_i m_i\right)\biggr| .
\end{equation}

\subsection{Redshift and diameter distance}

We consider a light ray which is emitted
from each shell toward an observer rest at the center. 
The light ray travels along a future 
directed ingoing radial null geodesic, where 
``ingoing'' refers to the direction 
from a shell
toward shells labeled by a smaller number. 

An ingoing radial null geodesic is specified by
a constant coordinate value $V$.
We denote the circumferential radius of the $i$th shell 
when it intersects the null geodesic as $R_{i}$.  
Labeling the outermost shell by $M$, 
the radius $R_{M}$ is equal to $x_{M}$(the initial radius
of the $M$th shell) and hence in the $M$th region, 
$V=V_{(-)M}(x_{M})$ is satisfied along the null geodesic. 
Thus, on the $(M-1)$th shell, the following relation holds:
\begin{equation}
  V_{(+)M-1}(R_{M-1})=V_{(-)M}(x_{M}).
\end{equation}
This equation determines $R_{M-1}$. 
By the same procedure, we obtain 
the circumferential radii of all the shells at the intersection 
with the null geodesic:
\begin{equation}
  V_{(+)i}(R_{i})=V_{(-)i+1}(R_{i+1}). \label{recurrent}
\end{equation}
We can determine $R_{i}$ from the given $R_{i+1}$
through this equation.

In order to derive the expression of redshift, we first write down the
components of the null geodesic tangent  
in the $i$th region, $k^{\mu}(i)$, which is given in 
the Kruskal null coordinate as  
\begin{equation}
  k^{U}(i)={r\over \mg}
             \exp\left({r\over 2 \mg}\right)\omega_{i},
\end{equation}
and the other components vanish, 
where $\omega_{i}$ is an integration constant associated 
with the geodesic equation. 
We require that the observed frequency
of the photon at each shell is uniquely determined.
The observed frequency, $\omega_{\rm ob}(i)$, 
at the $i$th shell is given by
\begin{eqnarray}
  \omega_{\rm ob}(i)
    &=&  -k_{\mu}(i)u_{(+)}^{\mu}(i)
     =   {1\over2}\omega_{i}{dV_{(+)i}\over d\tau} , 
    \label{eq:freq1}\\
  \omega_{\rm ob}(i+1) 
    &=&  -k_{\mu}(i)u_{(-)}^{\mu}(i+1)
     =   {1\over2}\omega_{i}{dV_{(-)i+1}\over d\tau}.
\label{eq:freq2}
\end{eqnarray}
Eqs. (\ref{eq:freq1}) and (\ref{eq:freq2}) gives the relation
between $\omega_{\rm ob}(i)$ and  
$\omega_{\rm ob}(i-1)$, for $i\geq2$, as 
\begin{equation}
\omega_{\rm ob}(i)=f(i)\omega_{\rm ob}(i-1) ,
\end{equation}
where
\begin{equation}
f(i) \equiv {dV_{(-)i}/d\tau \over dV_{(+)i-1}/d\tau}  .
\end{equation}
For the first region, a direct calculation leads to 
\begin{equation}
\omega_{\rm ob}(1)
  =  \left(   {dt_{(-)1}\over d\tau} + {dr_1\over d\tau} 
     \right) \omega_{\rm ob}(0)
  \equiv f(1) \omega_{\rm ob}(0),
\end{equation}
where $\omega_{\rm ob}(0)$ is the frequency of 
the light ray observed by an observer rest at the origin $r=0$. 
Thus, using the above relations, we obtain the 
redshift of the light ray emitted from the $i$th shell 
toward the observer rest at $r=0$ in the form
\begin{equation}
  1+z(i) \equiv {\omega_{\rm ob}(i)\over \omega_{\rm ob}(0)}
=\prod_{j=1}^{i}f(j) . \label{redshift}
\end{equation}

Our next task is to find the angular diameter-distance $\dA$ .
The definition of $\dA$ is
\begin{eqnarray}
  \dA \equiv {D\over \theta},
\end{eqnarray}
where $D$ is the physical length of the source perpendicular to the
line of sight, and $\theta$ is the observed angular size.
Since the space we are considering is spherically symmetric,
the diameter distance from the observer at the center
to the $i$th shell
agrees with the circumferential radius $R_i$ 
when the null geodesic intersects it;
\begin{eqnarray} 
  \dA (i) = R_i  . \label{diameter}
\end{eqnarray}

Now we can calculate $\dA$-$z$ relation in the dust-shell universe
using the relations
(\ref{recurrent}), (\ref{redshift}) and (\ref{diameter}).

\section{Results and Discussion}
\subsection{Setting the parameters of dust-shell model}
As mentioned in Sec. III the choice of 
the parameters and the initial hypersurface determines the 
behavior of a dust-shell universe.
Since we are interested in cases which have a FL limit,
we set the parameters so that they approach a FL model 
in the large $N$ limit.

We set the mass distribution of the
shells as
\begin{eqnarray}
\rho_i = \rho_{\rm c}\ {\rm (independent\ of}\ i{\rm )},
\label{eq:density} 
\end{eqnarray}
and $k_i$ in Eq. (\ref{expa}) as
\begin{eqnarray}
  k_i = k_{\rm c} .
\end{eqnarray}
Using $\rho_{\rm c}$ and $k_{\rm c}$, we define 
parameters $\Hsi$ and $r_{\rm H}$ by
\begin{eqnarray}\label{Hubbleconst}
  \Hsi^2 \equiv r_{\rm H}^{-2}\equiv
       {8\pi \over 3}\rho_{\rm c} - k_{\rm c} .
\end{eqnarray}
Further, we define
\begin{eqnarray}
  \Omega \equiv {8\pi \over 3}\rho_{\rm c}/\Hsi^2 .
\end{eqnarray}
Then,
\begin{eqnarray}
  k_{\rm c} = (\Omega - 1)\Hsi^2 .
\end{eqnarray}
In terms of FL models, $\Hsi$, $r_{\rm H}$ and $\Omega$ may be regarded as
the ``Hubble constant'', ``Hubble horizon radius'',
and the ``density parameter.''

For $x_i$, we put 
\begin{eqnarray}\label{eq:const-x}
   x_i(\tau_{\rm init}) = i\Delta x ,
\end{eqnarray}
with a constant interval $\Delta x$
\begin{equation}
\Delta x \equiv {\rH \over N_{\rm H}},
\end{equation}
where $N_{\rm H}$ is some positive integer.

Before we proceed, we estimate the magnitude of the radiation-like term in
Eq. (\ref{expa}). From 
Eqs. (\ref{eq:gmass-def}) and (\ref{eq:g-mass}) with
$m_{{\rm g}1}=0$, we find 
\begin{eqnarray}
m_{{\rm g}2n-1}&=&{1\over \NH^{3}}(n-1)^{2}(4n-1)\Omega\rH, 
\label{eq:g-mass-odd} \\
m_{{\rm g}2n}&=&{1\over \NH^{3}}n^{2}(4n-3)\Omega\rH,
\label{eq:g-mass-even}
\end{eqnarray}
where $n$ is a positive integer. 
From Eqs. (\ref{eq:dmass-def}) and (\ref{eq:curvature}), 
we find $\ms (i)= (m_{{\rm g}i+1}-\mg) /E_i$ and thus obtain
\begin{eqnarray}
m_{\rm s}(2n-1)&=&{6n^{2}-6n+1\over \NH^{3}}
{\Omega\over \sqrt{1-k_i x_i^2}}\rH, \label{eq:c-mass-odd} \\
m_{\rm s}(2n)&=&{6n^{2}\over \NH^{3}}
{\Omega\over \sqrt{1-k_i x_i^2}}\rH. \label{eq:c-mass-even}
\end{eqnarray}
Thus, when we consider a large $N$ limit with 
fixing $x_{i}=\rH(i/\NH)$, the proper mass $Gm_{\rm s}(i)={\cal
  O}(\NH^{-1})$  
is regarded as a small quantity, compared with $M_{+}(i)$ and 
 $m_{{\rm g} i}$. That is, the radiation-like term of Eq. (\ref{expa})
is of order $ \NH^{-2}$ of the other terms, 
and can be neglected when $\NH$ is sufficiently large.
We note that when 
$k_{\rm c}$ is positive (i.e., $\Omega>1$), 
the factor $\Omega / \sqrt{1-k_i x_i^2}$ in
Eqs.(\ref{eq:c-mass-odd}) and (\ref{eq:c-mass-even}) is 
larger than unity, and vice versa when $k_{\rm c}$ is negative.
This means that, in the dust-shell universe, 
 the effect of the local inhomogeneities 
(caused by condensing masses to a shell)
on the expansion rate appears
larger in a closed model and smaller in an open model
than that in a flat model
which has the same $\Hsi$ and $x_i$'s.
%
\subsection{Distance-redshift relation and averaged density}
As described in Sec.II, the choice of $e_i$ corresponds
to choosing an initial hypersurface. 
We fix the expansion law to be homogeneous, 
try some choices of $e_i$, and 
study which $e_i$ makes the distance-redshift relation behave 
like that of a FL model.
We will consider which physical meaning is carried by that choice,
especially in terms of averaged density.
We have already analyzed spatially flat cases (i.e., $k_c =0$)
in Paper I,
and found that the distance-redshift relation agrees with that of a 
(spatially flat) FL model quite well when we 
choose $e_i$ so that the averaged density is homogeneous.
The averaged density inside the $i$th shell $\bar{\rho}(i)$
is defined by dividing the ``mass'' contained within $r_i$
by the three-volume ${\rm Vol}(i)$ on the hypersurface up to $r_i$.
We may call the density field homogeneous
when $\bar{\rho}(i)$ is independent of $i$.

We here write down explicitly the three-volume 
\begin{eqnarray}
 {\rm Vol}(i)
&=&4\pi\int_{r_{i-1}}^{r_{i}}{r^{5\over2}\over
 \sqrt{(1+e_{i}^{2})r-2\mg}}dr \nonumber \\
&=&4\pi
\left[\sqrt{(1+e_{i}^2)r-{2\mg}}\left({r^{5/2}\over
  3(1+e_{i}^2)}+{5r^{3/2}{\mg}\over 6(1+e_{i}^2)^2} 
+{5{r}^{1/2}{\mg}^2\over 2(1+e_{i}^2)^3}\right)\right]^{r_i}_{r_{i-1}}\\
&+&
{20\pi {\mg}^3 \over (1+e_{i}^2)^{7/2} }
\ln \left({
\sqrt{(1+e_{i}^2)r_i-{2\mg}} + \sqrt{(1+e_{i}^2)r_i}
\over
\sqrt{(1+e_{i}^2)r_{i-1}-{2\mg}} + \sqrt{(1+e_{i}^2)r_{i-1}}
}\right) .
\end{eqnarray}
For $i=1$, ${\rm Vol}(1)$ is equal to $4\pi r_{1}^{3}/3$.

Let us study homogeneous density cases,
as in Paper I.
Now we have two kinds of mass in defining the averaged density;
proper mass $m_{\rm s}(i)$ defined in Eq. (\ref{eq:rmass-def})
and gravitational mass $\mg$ which determines the dynamics of its
outer shells.

We first take the hypersurface
so that $\bar{\rho}(i)$ 
using the gravitational mass is  
homogeneous.
We can define 
the gravitational mass of the $j$th shell to be
the difference between the gravitational mass parameters 
of the neighboring regions:
$m_{{\rm g}j+1} - m_{{\rm g}j} ~[= E_j \ms (j)]$. 
We add the gravitational mass of each shell up to $i-1$,
and add only half of the mass of the $i$th shell.
The averaged density thus defined is given as
\begin{equation}
  \bar{\rho}(i) \equiv
      \left\{\half E_i \ms (i)+\sum_{j\le i-1} E_j \ms (j)\right\}
      {\Big /}\sum_{j\le i}{\rm Vol}(j) 
   = \half (\mg +m_{{\rm g}i+1}){\Big /}\sum_{j\le i}{\rm Vol}(j) .
\end{equation}
As one can see from the definition of $\rho_{\rm c}$,
the choice of ${\rm Vol}(i)$ (and hence the choice of $e_i$)
which makes this averaged density homogeneous, i.e.,   
independent of $i$ (and equal to $\rho_{\rm c}$) is
\begin{eqnarray}
  {\rm Vol}(i) = {4\pi\over 3}(r_i^3 - r_{i-1}^3 ) .
\end{eqnarray}
We plot the $\dA$-$z$ relation of the dust-shell models 
fixed in this way in Figs. 2-4
(choice A).  The employed parameters are $\Omega = 1.0,0.9$ and $1.1$,
$\Hsi = 1.0$, and 
the number of shells within the initial Hubble radius $\NH$
is set to be $4$ and $10$.
The outermost shell $M$ from which a light is emitted 
is chosen to be $2.5 \times \NH$.
We also plot the $\dA$-$z$ relation in FL models,
which is determined by 
the initial Hubble parameter $H_{\rm I}$, initial density parameter
$\Omega_{\rm I}$,
and the redshift of the initial time slice $z_{\rm I}$.
These parameters are set using the parameters of the 
corresponding shell model as
 $H_{\rm I} = \Hsi$, $\Omega_{\rm I} = \Omega$, 
and $z_{\rm I}$ equal to the redshift of the outermost shell
for $\NH = 10$.
In the flat case, the data of the dust-shell models 
 agree with the FL relation as seen in Paper I.
In the non-flat cases, however, the FL models chosen in this way
do not approximate the dust-shell models.
Even when we increase the number density of the shell,
no improvement is obtained.
Thus, using the gravitational mass of the shell in averaged density 
is inappropriate.

Next, we try the choice using the proper mass 
in defining the averaged density,
\begin{equation}
  \bar{\rho}(i) \equiv
      \left\{\half m_{\rm s}(i)+\sum_{j\le i-1} m_{\rm s}(j)\right\}
      {\Big /}\sum_{j\le i}{\rm Vol}(j) .
\end{equation}
After some manipulation, 
one finds that the volume element should satisfy the relation
\begin{eqnarray}
 {\rm Vol}(i)  = 
{4\pi\over 3}\left( {r_i^3\over E_i}-{r_{i-1}^3\over E_{i-1}}\right)
+\left( {1\over E_{i-1}}-{1\over E_{i}}\right) \mg ,
\end{eqnarray}
in order to make the averaged density 
$\bar{\rho}(i)$ equal to $\rho_{\rm c}$ (choice B). 
In a flat model, this choice is the same as choice A.
The $\dA$--$z$ relations for open and closed models are plotted in
Figs. 5 and 6. 
These plots show that the relations agree quite well with FL models.
Thus, we should use the proper mass in defining the averaged density.
We note that 
owing to the amplification of inhomogeneity
which appeared in Eqs.(\ref{eq:c-mass-odd}) and (\ref{eq:c-mass-even}), 
closed universe shows slight deviation
one can notice when compared with the other cases.

For comparison,
we try the orthogonal slice,
since, in a FL model, the simultaneous hypersurface is
orthogonal to the trajectory of matter (choice C).
We choose $e_i$ so that the vector $\ell^a$ is
orthogonal to the trajectory of each shell. 
From the condition $\ell_{a} u^{a}_{(+)}(i-1) =0$ at $r=r_{i-1}$,
we obtain $e_i = d r_{i-1}/d\tau$.
Fig. 7 shows the $\dA$-$z$ relation in this choice\footnote{
We only displayed an open model. 
There is no solution when $\Omega >1$ which satisfies 
the conditions (\ref{eq:const-x}) and $\ell_{a} u^{a}_{(+)}(i-1) =0$
simultaneously for $\NH =10,M = 25$.}.
We can see a mild deviation from the FL model.
\subsection{Discussion on the dust-shell model}
In paper I we concluded that the $\dA$-$z$
relation 
in a dust-shell universe behaves like a flat FL universe,
when the following conditions are satisfied:
the expansion law resembles the flat FL model;
the behavior of averaged density field is scale-independent
when we increase the scale of averaging;
the averaged density agrees with $\rho_{\rm c}$
(defined by Eq.[\ref{eq:density}]).
The discreteness nature of the shell model
plays no significant role in our spherical model.

However, our analysis was limited to spatially-flat cases 
(Einstein de-Sitter model). 
In this paper, we have examined the non-flat cases, and
confirmed that the above statement remains valid.
Moreover, paper I could not answer
which mass should be used in defining the average
density, proper mass which appears in the energy-momentum
of dust shell,
or gravitational mass which specifies the parameter in
Schwarzschild spacetime; 
the latter includes the gravitational potential energy 
and the kinetic energy of the shell.
In the spatially flat models, they balance and the two masses agree
with each other.
From the results of this paper,
now it becomes clear that the proper mass should be used in defining
the averaged density. 
This might at first sound strange since the geometry of 
each region is 
determined by the gravitational mass which includes the kinetic energy
and the potential energy. 
Where has their information gone?
Reexamining the Hubble equation of FL models and
the expansion law of the dust-shell models,
one notices that it is contained in the curvature term.
In Eq. (\ref{eq:curvature}), the curvature term is expressed as the ratio of
the proper mass to the 
difference of gravitational mass between the inside and 
the outside regions. 
Thus, the curvature gives the ratio of the total energy to the proper
mass. 

In our model, 
the expansion law for the circumferential radius
is homogeneous when there are enough number of shells;
the effect on the expansion rate of density inhomogeneities 
is small (of order $\NH^{-2}$).
Fixing the expansion law like FL,
we studied the connection between the 
$\dA$-$z$ relation and the averaged density.
We have found 
that if we make the averaged density (which is defined using 
the three-volume element of the hypersurface and the proper mass)
homogeneous,  
the $\dA$-$z$ relation agrees well with that of a FL model.
That is, 
there exists a strong connection among 
the homogeneous averaged density, the homogeneous expansion law,
and the FL like distance-redshift relation, 
in spite of the discrete nature 
(local inhomogeneity of matter distribution) of the dust-shell model.
This will support the ``averaging hypothesis'' that a universe
behaves like a FL model in spite of small-scale fluctuations
of density field,
if its averaged density field is homogeneous on large scales. 
This implies that even if there exist large wall-like structures,
our universe is approximated by a FL model on larger scales
if the walls satisfy the above conditions.

%
Now we give some comments on the small $N$ cases;
the cases where the radiation-like term in the expansion law
is not negligible.
One may expect that a 
FL model including radiation term can fit them, 
but this does not work.
It should be noticed that the radiation-like term is inhomogeneous
(i.e., dependent on $i$) in the cases considered in this 
paper.
Then, there is no reason one may expect that the behavior of observables
obeys a FL like relation;
it is natural that we cannot fit them with
FL models including radiation term.
That is, it is impossible to approximate such an inhomogeneous
model that has a significant large-scale inhomogeneity by a FL model.

\subsection{Implication on the universe filled with bound objects}
We can obtain further insight into the treatment of 
inhomogeneous universes by investigating the dust-shell universe
from a different viewpoint,
especially in connection with gravitationally bound objects.
We discuss its implications on the 
estimation of the cosmological parameters 
of a locally inhomogeneous universe.

We start by considering the construction of a dust 
shell from small particles.
In Appendix, we consider momentarily static initial data
in which the intrinsic 
metric of the spacelike hypersurface is conformally flat and there are 
arbitrary number of compact objects. 
By construction of the
solution, locations of the objects are arbitrary.
Hence, by arranging an infinite number of infinitesimal objects
with an appropriate procedure,
we can construct 
an infinitesimally thin shell which is momentarily static.
As discussed in Appendix, a spherical shell constructed by this procedure 
is likely to be regarded as a dust shell treated in this paper.
Hence, by arranging sufficient number of such shells
by the manner shown in this paper, 
we can obtain a system of compact objects which well imitates a FL universe. 

Next, 
let us consider a closed FL universe filled with baryonic dust fluid
and a dust-shell universe (or a universe filled with compact 
objects) which have the same baryonic mass.
The analysis of the dust-shell universe implies that 
the sum of the proper mass of all the shells in the closed 
dust-shell universe is the same as the baryonic 
mass of the closed FL universe,
when the distance-redshift relations of these universes well agree with 
each other.
On the other hand, 
by the investigation in Appendix, we find that the proper mass 
of a shell is the sum of the gravitational 
mass of the objects composing the shell. 
Here we should recall that the 
gravitational mass of an object is in general different from its baryonic
mass; there is a gravitational mass defect and the difference 
between the gravitational mass and baryonic mass is recognized to be
the binding energy of the object. 
Thus, if the shells are composed of highly
bound objects, the sum of the gravitational mass 
of all the objects is much smaller than the baryonic mass.
Therefore, even if the total baryonic mass 
of the universe composed of gravitationally bound objects
is the same as that of 
the FL universe filled with 
the baryonic dust fluid, 
the distance-redshift relations of these universes 
can highly disagree with each other.

Now let us study the difference between these universes with the same 
amount of baryonic mass quantitatively.
The line element of 
the closed FL universe filled with the dust fluid
is written as
\begin{equation}
ds^{2}=-d\tau^{2}+a^{2}(\tau)\left(d\chi^{2}+\sin^{2}\chi
  d\Omega^{2}\right).
\end{equation}
The solution of the Einstein equation can be written in the form
\begin{eqnarray}
a&=&{2M\over 3\pi}\left(1-\cos\eta\right), \label{eq:s-factor}\\
\tau&=&{2M\over 3\pi}\left(\eta-\sin\eta\right)\label{eq:c-time}, 
\end{eqnarray}
where $M$ is the total mass of the dust and $0 \leq \eta \leq 2\pi$. 
In the dust-shell model, 
the mass $M$ is given by the sum of the gravitational mass 
of all the objects in the universe
if the shells are composed of bound objects;
we denote it by $M_{\rm G}$. 
On the other hand, in the FL universe filled with 
the baryonic dust fluid, $M$ is the total baryonic mass $M_{\rm B}$. 
When the two universes have the same amount of baryons,
$M_{\rm G}< M_{\rm B}$ holds. 

Here we study how the Hubble parameter 
$H$ and density parameter $\Omega$ 
 at a fixed age $\tau=\tau_{0}$ change
when we change the mass parameter $M$.
The Hubble parameter and the density parameter 
are defined by 
\begin{eqnarray}
H&\equiv& {1\over a}{da\over d\tau}
={3\pi\sin\eta \over 2M(1-\cos\eta)^{2}}, \label{eq:Hubble} \\
\Omega&\equiv&{4M\over 3\pi a^{3}H^{2}}
={2(1-\cos\eta)\over \sin^{2}\eta},\label{eq:D-parameter}
\end{eqnarray}
where $M$ and $\eta$ is connected through the relation
$\tau_0 = {2M\over 3\pi}\left(\eta-\sin\eta\right)$.
Then the derivative of $M$ with respect to $\eta$ is given by
\begin{equation}
\left.{\partial M\over \partial \eta}\right|_{\tau_0}
=-{3\pi\tau_{0}(1-\cos\eta)\over
2(\eta-\sin\eta)^{2}}. \label{eq:M-dif}
\end{equation}
From Eqs.(\ref{eq:Hubble}), (\ref{eq:D-parameter}) 
and (\ref{eq:M-dif}), we obtain
\begin{eqnarray}
\left.{\partial H\over \partial M}\right|_{\tau_0}
&=&{3\pi\left\{3\left(\eta-\sin\eta\right)
-\eta\left(1-\cos\eta\right)\right\}\over 2M^{2}(1-\cos\eta)^{3}}, \\
\left.{\partial \Omega\over \partial M}\right|_{\tau_0}
&=&-{2(1-\cos\eta)(\eta-\sin\eta)\over M\sin^{3}\eta}.
\end{eqnarray}
One can confirm the positivity of $\partial H/ \partial M$ and  
hence the Hubble parameter increases with increasing mass $M$. 
On the other hand, the density parameter $\Omega$ is 
a decreasing function of $M$ in the expanding phase while 
an increasing function in the contracting phase. 
This implies that if the dust shells are made of bound objects, 
the Hubble parameter of the dust-shell universe 
is smaller, and the density parameter is larger, than 
those of the FL universe filled with baryonic dust fluid 
(in the expanding phase) with the same amount of baryons 
and the same age, since the relation $M_{\rm G}< M_{\rm B}$ holds. 

In the limit of $\tau/M\rightarrow0$
(accordingly $\eta\rightarrow0$), the closed FL universe filled 
with dust fluid behaves like the Einstein-de Sitter universe.
The behavior of $H^{-1}\partial H/ \partial M$ in this limit 
is easily obtained:  
\begin{equation}
\left.{1\over H}{\partial H\over \partial M}\right|_{\tau_0}
\longrightarrow {1\over 30M}~~~~~{\rm for}~~~~
{\tau_{0}\over M} \longrightarrow 0.
\end{equation}
Using the above equation, the difference in the Hubble parameter
of the dust-shell universe, $H_{\rm DS}$,
and that of the FL universe 
filled with the baryonic dust fluid, $H_{\rm FL}$,
is given by
\begin{equation}
{H_{\rm FL}-H_{\rm DS}\over H_{\rm FL}}\sim 
{1\over 30M_{\rm B}}(M_{\rm B}-M_{\rm G}) < {1\over 30}.
\end{equation}
Thus, at the nearly Einstein-de Sitter stage, the effect of the 
mass defect of the objects in the universe is rather small. 
(The difference in the density parameters of these universes 
vanishes in this limit.)
On the other hand, at the stage of the maximum expansion $\eta=\pi$, 
the derivatives, $H^{-1}{\partial H/ \partial M}$ and
$\partial \Omega/ \partial M$, blow up. 
This is simply because $H$ goes to $0$ at $\eta=\pi$ and thus
the diverging behavior itself has no serious consequence.
However, this indicates the tendency that
the effect of the binding energy of the compact objects 
becomes somewhat larger than the Einstein-de Sitter epoch
when the curvature of the universe is 
comparable to the energy density of the dust fluid. 

The observation of CMBR strongly suggests that our universe was 
highly isotropic and homogeneous at least on the last scattering 
surface. Hence in the study of the universe in the early stages, 
the linear perturbation analysis is powerful. 
In order to perform the linear analysis, we need to fix the background 
FL universe,
whose cosmological parameters
are usually determined by the observation of our neighborhood. 
The universe observed today is, however,  
highly inhomogeneous and the inhomogeneities  
may prevent the correct determination of the background FL universe. 
The above discussion implies that 
taking account of the binding energy may be important 
in estimating the density parameter and the Hubble parameter
near the maximum expansion,
if the universe is filled with highly bound objects. 
On the other hand, when the universe is in the stage 
when the curvature of the universe is not dominant, 
the effect of the mass defect is rather small. 
We note that, however, 
in order to discuss the physical quantities of the inhomogeneous universe
(e.g., the age of the universe) and its whole time evolution
using a FL model constructed by the nearby observations,
we have to know the behavior of the scale factor in the transition 
regime from the almost FL universe to the inhomogeneous one.  
This problem is now under investigation and   
will be given elsewhere\cite{rf:SN}. 

\section{Summary}
We studied the behavior of $\dA$-$z$ relation in a
spherically symmetric dust-shell universe
where the mass distribution is discrete.
Extending the treatment in our previous paper
which only spatially flat models were considered,
we analyzed non-flat cases.
We compared the distance-redshift relation of dust-shell universe 
with that of FL models.
In particular, we examined
the behavior of the averaged density of the dust-shell universe 
when the two $\dA$-$z$ relations are similar.
We found that the $\dA$-$z$ relation observed at the center
agrees quite well with that of a flat FL model 
if the following conditions are satisfied:
(i) the expansion law of the circumferential radius of the shells
resembles the Hubble equation of a spatially flat FL model,  
(ii) the behavior of averaged density around the observer
at the center is scale-independent as we increase the scale on
which we take the average, and
(iii) the averaged density agrees with 
the energy density of the FL model.
In defining the averaged density,
we take the total proper mass of the shells and divide it 
by the three-volume of the initial hypersurface.
We noted that the choice of the initial hypersurface
relates the expansion law to the averaged density.

The effect of discreteness of mass distribution appears 
in the equation of motion of each dust shell. This effect becomes
smaller as we increase the number density of shell,
though we found that the positive curvature has tendency to enhance the
inhomogeneity effect. 
We conclude that, in this spherical dust-shell model,
the discrete nature of matter distribution 
plays no significant role 
in discussing the observed quantities such as $\dA$ and $z$,
as long as the expansion law and the averaged density field is
sufficiently homogeneous in the sense described above. 
This supports the averaging hypothesis that a universe
is described by a FL model if the universe is homogeneous 
when the density is averaged on a large scale than the scale 
of the inhomogeneities. 
This may also imply that even if there exist large wall-like structures,
our universe is approximated by a FL model on larger scales
if the walls satisfy the above conditions.

However, we have to keep in mind that our model is 
highly idealized and our analysis is limited only to 
spherically symmetric cases.
In general, local inhomogeneities strongly affect 
the light propagation, giving rise to dispersions
in the observed $\dA$-$z$ relation\cite{rf:KANTOWSKI,rf:DYERROEDER1,rf:SEF}.
It will be also interesting to study cases when the light ray 
enters a shell in a non-radial direction.

It still remains unclear whether 
three-dimensional discreteness 
have a significant effect on the dynamics of the universe. 
The study of the momentarily static initial data in Appendix
strongly 
suggests that when the universe is filled with bound
objects, the dynamics of the universe is determined by the 
gravitational mass density of the objects but not by the 
baryonic mass density. The gravitational mass of a compact object
is in general different from its baryonic mass due to 
the gravitational mass defect. 
The effect of the mass defect on the dynamics of the 
universe is significant at the stage of the curvature-dominant phase 
while it is not significant at the early stage of the universe, i.e., 
the stage during which the universe behaves as the Einstein-de Sitter 
universe. However, in general, there are both the bound and 
unbound objects. Hence we should consider a situation including 
both objects and investigate their effects on the dynamics, which is 
left for our future work. 

\section*{Acknowledgements}
We would like to thank H.~Sato for encouragement.
N.S. and T.H. are supported by Japan Society for 
the Promotion of Science for Young Scientists Grant Nos. 3167 and 
9204. This work is partially supported by
Grant-in-Aid for Creative Basic Research(09NP0801) 
and for Scientific Research B(09440106) from the Japanese
Ministry of Education, Science, Sports and Culture.

\appendix
\section*{Momentarily Static Initial Data of a Spherical Shell 
Composed of Bound Objects}

In order to get an insight into the origin of the proper mass 
$m_{\rm s}$ of a spherical shell, 
we consider momentarily static initial data of which the extrinsic 
curvature vanishes. 
The intrinsic metric of the three-dimensional spacelike 
hypersurface is  assumed to 
be conformally flat,
\begin{equation}\label{eq:conf}
dl^{2}=\psi^{4}({\vec x})d{\vec x}^{2},
\end{equation}
where $\vec x$ is a position vector.  
Then the Hamiltonian constraint is written as
\begin{equation}
\Delta\psi=-2\pi\rho_{{\rm H}}\psi^{5},
\end{equation}
where $\Delta$ is the Laplacian operator in the flat space and 
$\rho_{\rm H}$ is the energy density for an observer 
whose trajectory is normal to the spacelike hypersurface. 

We introduce the gravitational mass density defined by
\begin{equation}
\rho_{\rm G}\equiv \rho_{\rm H}\psi^{5}.
\end{equation}
Let us consider a situation where $n$ ``spherical'' objects exist. 
We give $\rho_{\rm G}$ by
\begin{equation}
\rho_{\rm G}=\sum_{I=1}^{n}\rho_{{\rm G}I},
\end{equation}
where 
\begin{equation}
\rho_{{\rm G}I}=\rho_{{\rm G}I}\left(|{\vec x}-{\vec x}_{I}|\right)\geq0,
\end{equation}
for $|{\vec x}-{\vec x}_{I}|\leq\ell_{I}$ (radius 
of the $I$th object),  
while $\rho_{{\rm G}I}$
vanishes for $|{\vec x}-{\vec x}_{I}|> \ell_{I}$.
The solution of the Hamiltonian constraint is then 
written in the form
\begin{equation}
\psi=1+\sum_{I=1}^{n}\psi_{I},  
\end{equation}
where $\psi_{I}$ satisfies the equation
\begin{equation}
\Delta \psi_{I}=-2\pi\rho_{{\rm G}I}, \label{eq:p-eq}
\end{equation}
with the boundary condition, $\psi_{I}\rightarrow 0$  
for $|{\vec x}-{\vec x}_{I}|\rightarrow \infty$. 

Integrating Eq. (\ref{eq:p-eq}), the solution of 
the Hamiltonian constraint is easily obtained:
\begin{equation}
\psi=1+2\pi\sum_{I=1}^{n}\int_{|{\vec x}-{\vec x}_{I}|}^{\infty}
dyy^{-2}\int_{0}^{y}dx x^{2}\rho_{{\rm G}I}(x).\label{eq:Sol1}
\end{equation}
For the vacuum region, the above solution takes a simple form
\begin{equation}
\psi=1+{1\over2}\sum_{I=1}^{n}{m_{I}\over |{\vec x}-{\vec x}_{I}|},
\label{eq:Sol2}
\end{equation}
where the parameter $m_{I}$ is defined by
\begin{equation}
m_{I}\equiv 4\pi \int_{0}^{\ell_{I}}dxx^{2}\rho_{{\rm G}I}(x).
\end{equation}
We also consider the proper mass $m_{{\rm p}I}$ of 
the $I$th object defined by
\begin{equation}
m_{{\rm p}I}\equiv \int d^{3}x\rho_{{\rm H}I}\psi^{6}
=\int d^{3}x\rho_{{\rm G}I}\psi.
\end{equation}
If the compact objects are composed of dust fluid, $m_{{\rm p}I}$ is
the conserved  
rest mass (baryonic mass).
The above integral is easily performed to give
\begin{equation}
m_{{\rm p}I}=m_{I}\left(1+\delta_{I}
+{1\over2}\sum_{J\neq I}{m_{J}\over |{\vec x}_{I}-{\vec x}_{J}|}\right),
\end{equation}
where
\begin{equation}
\delta_{I}\equiv {1\over m_{I}}\int d^{3}x \rho_{{\rm G}I}\psi_{I}=
{8\pi^{2}\over m_{I}}
\int_{0}^{\ell_{I}}dz z^2 \rho_{{\rm G}I}(z)\int_{z}^{\infty}dyy^{-2}
\int_{0}^{y}dxx^{2}\rho_{{\rm G}I}(x).
\end{equation}

Now let us consider the ``gravitational mass'' of the $I$th object,  
which includes the gravitational binding energy. 
In order to obtain it, we replace the compact object
by an Einstein-Rosen bridge with the 
same mass parameter $m_{I}$. 
This means that we employ solution (\ref{eq:Sol2}) 
even inside the object. 
Now instead of the compact object, we have a ``sheet'' with an 
asymptotic region $|{\vec x}-{\vec x}_{I}|\rightarrow 0$. 
In this asymptotic region, one can define the ADM mass 
(gravitational mass) 
$M_{I}$ which corresponds to the total energy of the $I$th object.
Brill and Lindquist\cite{rf:BRILL} showed that $M_{I}$ is given by
\begin{equation}\label{ADM-mass}
M_{I}=m_{I}\left(1+{1\over2}\sum_{J\neq I}^{n} 
{m_{J}\over |{\vec x}_{I}-{\vec x}_{J}|}\right).
\end{equation}
Then the gravitational binding energy $E^{I}_{\rm bind}$ of 
the $I$th object is naturally defined as
\begin{equation}
E^{I}_{\rm bind}=M_{I}-m_{{\rm p}I}=-m_{I}\delta_{I}. \label{eq:Bind-E}
\end{equation}
The above equation implies that $\delta_{I}$ is the specific binding 
energy of the $I$th object. 

Here let us consider a situation in which compact objects with almost 
identical mass $m_{I}\sim m$ are distributed 
homogeneously on spheres with a common center labeled by $i=1,...,N$.
The $i$th sphere contains the objects labeled by $I$ 
in the range $n_{i}+1\leq I \leq n_{i+1}$,
where $n_{1}=0$, $n_{i}<n_{i+1}$, and $n_{N+1}$ is equal to the 
total number of the objects $n$.
Note that the position vector of the $I$th object on the $i$th sphere
satisfies $|{\vec x}_{I}|=R_{i}$.
Let us denote the sum of the parameter $m_I$ inside the
$i$th region as
\begin{equation}
M_{{\rm g}i} \equiv \sum_{I=1}^{n_{i}}m_{I}.
\end{equation}
We consider a limit where the number of the objects on each shell
goes to infinity with $M_{{\rm g}i}$ fixed:
$n_{i+1}-n_{i}\rightarrow \infty$ for all $i$'s. 
By this procedure, we obtain a system of 
infinitely thin $N$ spherical shells with a common center. 
This system has the same configuration as that treated in this paper 
(see Fig.1).  
In the limit where $n_{i+1}-n_{i}$ is very large,
the mean separation, $L$, between the objects is given by
\begin{equation}
L\sim \sqrt{4\pi r_{i}^{2}\over n_{i+1}-n_{i}},
\end{equation} 
while the mass parameter $m_{I}$ is 
\begin{equation}
m_{I}\sim {M_{{\rm g}i+1}-M_{{\rm g}i}\over n_{i+1}-n_{i}}. 
\end{equation}
We assume that the radius of each object
is $\ell_{I}=\alpha m_{I}$, where $\alpha$ is a constant. 
If we consider an Einstein-Rosen bridge instead of the object, 
we choose the constant $\alpha$ to be $1/2$. 
This assumption implies that in the large-$(n_{i+1}-n_{i})$ limit, 
$\ell_{I}/L$ approaches to zero since it is proportional 
to $(n_{i+1}-n_{i})^{-1/2}$. 
Hence a shell obtained by this limit is extremely sparse. 
In order to know whether the shell obtained by this limiting procedure 
is a dust shell, we need to study the time evolution
of this initial data. Although we do not investigate its time 
evolution here, 
it seems likely that a shell composed of infinite number of 
infinitesimally small objects is a dust shell
by the following reason.
Since the mean separation  
between the objects is infinitely larger than the radius of each object,
a direct collision between the objects is impossible
when the shell is shrinking.
Further, since $m_{I}/L$ is extremely smaller than unity, 
the effect of the gravity of a nearby particle
which causes a non-radial motion is likely to be very small.
These two should work to keep the spherical symmetry 
and the radial motion of the shell during its evolution.

In order to obtain the relation between $M_{{\rm g}i}$ and 
the mass parameter $m_{{\rm g}i}$ of the Schwarzschild spacetime 
in the $i$th region, we investigate the line element in 
the $i$th region.
The conformal factor in Eq. (\ref{eq:conf}) in the $i$th region 
is written as
\begin{equation}
\psi=A_{i}+{M_{{\rm g}i}\over 2R}, \label{eq:cfactor}
\end{equation}
where $A_{i}$ is a constant and $R\equiv |{\vec x}|$. 
From the continuity of the conformal 
factor, we obtain a recurrent relation for $A_{i}$ as
\begin{equation}
A_{i}=A_{i+1}+{1\over 2R_{i}}\left(M_{{\rm g}i+1}-M_{{\rm g}i}\right).
\label{eq:A-eq}
\end{equation}
Since the value of $A_{N+1}$ in the outermost region 
is unity, we can obtain $A_{i}$ for 
$1\leq i\leq N$ by the above equation. 
Eq. (\ref{eq:cfactor}) gives the line element as
\begin{equation}
dl^{2}=A_{i}^{4}\left(1+{M_{{\rm g}i}\over 2A_{i}R}\right)^{4}
\left(dR^{2}+R^{2}d\Omega^{2}\right).
\end{equation}
Introducing a new radial coordinate $X=A_{i}^{2}R$, 
the above line element becomes
\begin{equation}
dl^{2}=\left(1+{A_{i}M_{{\rm g}i}\over 2X}\right)^{4}
\left(dX^{2}+X^{2}d\Omega^{2}\right),
\end{equation}
which is the line element of the Schwarzschild spacetime
in the isotropic coordinate. 
Then the mass parameter 
$m_{{\rm g}i}$ is now trivially obtained as
\begin{equation}
m_{{\rm g}i}=A_{i}M_{{\rm g}i}. \label{m-M-relation}
\end{equation}

As reviewed in Sec. II, the dynamics of an infinitely thin shell 
is treated by  Israel's formalism. 
The equation for the circumferential radius $r_{i}$ of 
the $i$th shell is given by Eq. (\ref{motion}): 
\begin{equation}
\left({dr_{i}\over d\tau}\right)^{2}
=\left({M_{-}(i)\over m_{\rm s}(i)}\right)^{2}-1+{2{\bar M}_{i}\over r_{i}}
+{m_{\rm s}^{2}(i)\over 4r_{i}^{2}}. \label{eq:shell-eq}
\end{equation}
The proper mass $m_{\rm s}(i)$ of a dust shell
is conserved during its time evolution.
We will write down $m_{\rm s}(i)$ using the 
mass parameters of the objects composing the shell,
by comparing the above equation with the solution of (\ref{eq:conf}).
The momentarily static situation corresponds to the moment of 
maximum expansion $dr_{i}/d\tau=0$. 
Hence from Eq. (\ref{eq:shell-eq}), 
we obtain the relation
\begin{equation}
\left({M_{-}(i)\over m_{\rm s}(i)}\right)^{2}-1+{2{\bar M}_{i}\over r_{i}}
+{m_{\rm s}^{2}(i)\over 4r_{i}^{2}}=0. \label{eq:ms-eq}
\end{equation}
In the present situation, we know the gravitational mass $m_{{\rm g}i}$ 
and the 
circumferential radius $r_{i}$ which is related with $R_{i}$ by
\begin{equation}
r_{i}=R_{i}\left(A_{i+1}+{M_{{\rm g}i+1}\over 2R_{i}}\right)^{2}.
\label{eq:r-R-relation}
\end{equation}
Hence Eq. (\ref{eq:ms-eq}) is regarded as an algebraic 
equation to determine the proper mass $m_{\rm s}(i)$ of 
the $i$th shell. 
The positive roots of this equation are given as
\begin{equation}
m_{\rm s}(i)=m_{\rm s\pm}=r_{i}\left(\sqrt{1-{2m_{{\rm g}i}\over r_{i}}}
\pm\sqrt{1-{2m_{{\rm g}i+1}\over r_{i}}}\right), 
\label{eq:ms-sol}
\end{equation}
where $m_{{\rm g}i}<m_{{\rm g}i+1}$ is assumed. 
In order to make the 
meanings of the above roots clear, we consider 
Eq. (\ref{eq:ex-time}). 
At the moment of the maximum expansion,  
$r_{i}$ should be larger than or equal to $2m_{{\rm g}i+1}$. 
First we consider the case when $r_{i}>2m_{{\rm g}i+1}$. 
Then we easily find that $dt_{(+)i}/d\tau$ is negative 
for $m_{{\rm s}}(i)=m_{{\rm s}+}$ and  positive
for $m_{{\rm s}}(i)=m_{{\rm s}-}$. 
Together with Eq. (\ref{eq:r-R-relation}), 
the positivity of $dt_{(+)i}/d\tau$ 
implies that $m_{{\rm s}}(i)$ should be equal to $m_{{\rm s}-}$ for
$R_{i}>m_{{\rm g}i+1}/2$
(i.e., the direction of the time coordinate $t_{(+)}$ should agree with
that of the proper time $\tau$ in this asymptotic region),
while it should be equal to $m_{{\rm s}-}$ for 
$R_{i}<m_{{\rm g}i+1}/2$.  
When $r_{i}=2m_{{\rm g}i+1}$, i.e., $R_{i}=m_{{\rm g}i+1}/2$, 
the solution $m_{{\rm s}+}$ agrees with $m_{{\rm s}-}$ and hence 
$m_{{\rm s}}(i)$ is equal to $m_{{\rm s}\pm}$ in this case. 

From Eqs.(\ref{eq:A-eq}) and (\ref{m-M-relation}), we find 
\begin{eqnarray}
m_{{\rm g}i}&=&\left\{A_{i+1}+{1\over 2R_{i}}\left(M_{{\rm g}i+1}-M_{{\rm g}i}\right)
  \right\}M_{{\rm g}i}, \\
m_{{\rm g}i+1}&=&A_{i+1}M_{{\rm g}i+1}.
\end{eqnarray}
Using Eqs.(\ref{eq:r-R-relation}),
(\ref{eq:ms-sol}) and the above equations, we obtain 
\begin{equation}
m_{\rm s}(i)=\left(M_{{\rm g}i+1}-M_{{\rm g}i}\right)
\left(A_{i+1}+{M_{{\rm g}i+1}\over 2R_{i}}\right),
\end{equation}
imposing $dt_{(+)i}/d\tau >0$.
On the other hand, in the spherical-shell limit, 
the gravitational mass $M_{I}$ in Eq. (\ref{ADM-mass})
of the $I$th object on the $i$th shell
becomes 
\begin{equation}
M_{I}\longrightarrow m_{I}\left(A_{i+1}
+{M_{{\rm g}i+1}\over 2R_{i}}\right).
\label{eq:g-massa}
\end{equation}
Hence we find in the limit $n_{i+1}-n_{i}\rightarrow\infty$ 
for all the shells, 
\begin{equation}
\sum_{I=n_{i}+1}^{n_{i+1}}M_{I} \longrightarrow m_{\rm s}(i). 
\label{eq:sg-relation}
\end{equation}
This equation implies that the proper mass of the shell 
is the sum of the gravitational mass of each object, 
not the sum of the proper mass of each object.

\newpage
\begin{figure}
\epsfysize=5cm \epsfbox{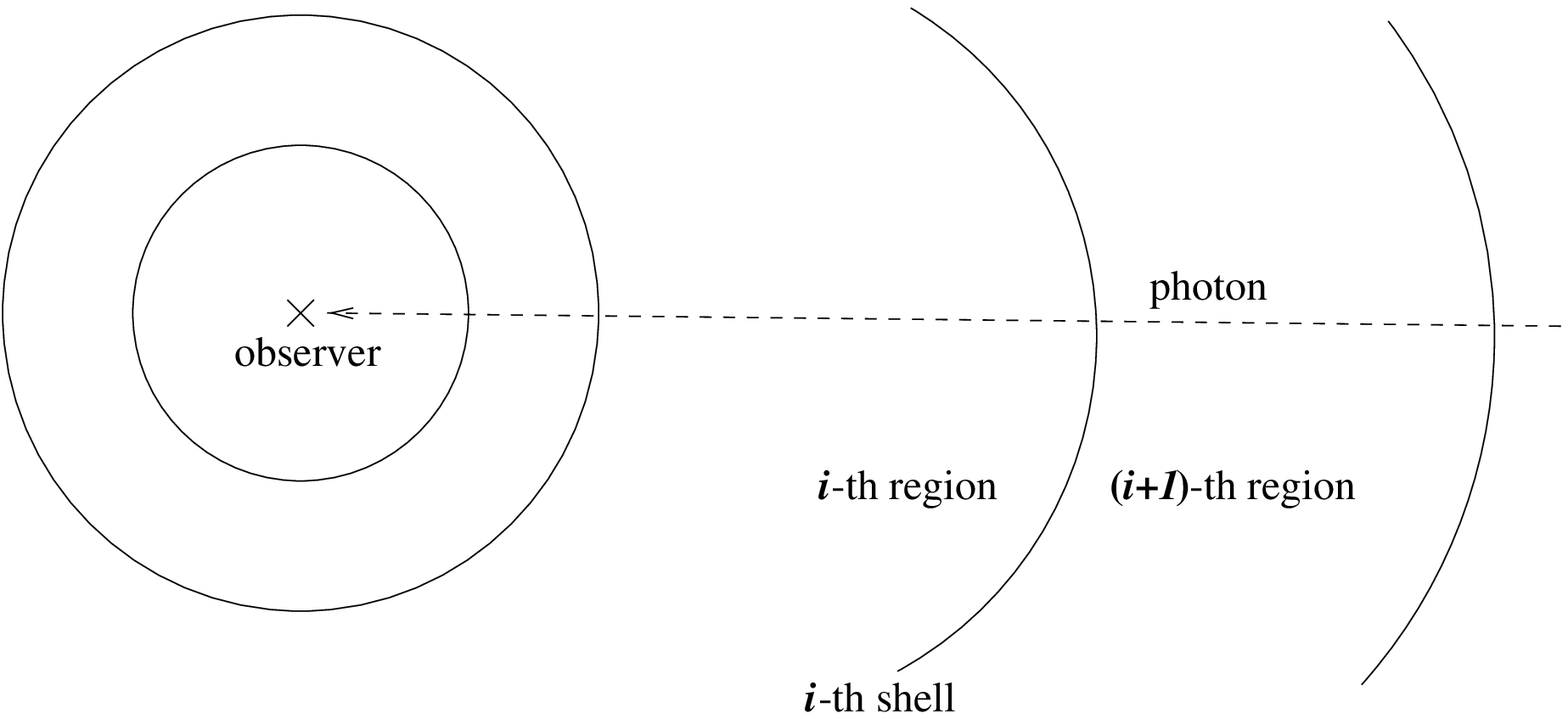}
\caption{Dust-shell universe. Between the shells the space is empty.
}
\end{figure}%
\begin{figure}[ht]
\epsfysize=8cm \epsfbox{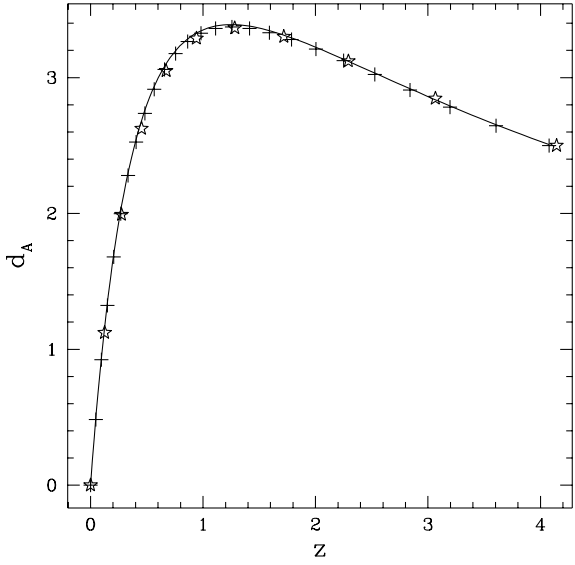}
\caption{Angular diameter distance-redshift relation in 
spatially flat ($k_c = 0$) dust-shell universe
for choice A.
Data points are shown by cross ($+$) and star($\star$).
The number of shells within the initial Hubble horizon $\NH$ 
is $4(\star)$ and $10(+)$.
The total number of shells $M$ is taken to be $2.5\times\NH$.
The solid line 
shows the $\dA$-$z$ relation in a flat
FL model with $H_\protect{\rm i\protect} = \Hsi$.
The redshift of the initial hypersurface is identified with the
redshift of the outermost shell for the case $\NT =25$, i.e.,
$z_\protect{\rm I\protect} = z(i=25)$. 
As shown in Paper I, the deviation between the flat FL model and the
dust-shell universe is almost unrecognizable.
}
\end{figure}
\begin{figure}[t]
\epsfysize=8cm \epsfbox{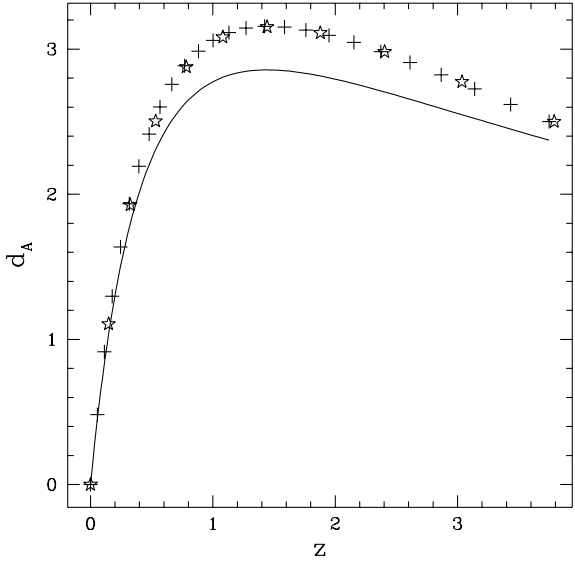}
\caption{Same with Fig.1 for open models ($\Omega = 0.9$).
The solid line 
shows the $\dA$-$z$ relation in an open 
FL model with $\Omega = 0.9, H_\protect{\rm I\protect} = \Hsi$,
$z_\protect{\rm I\protect} = z(i=25)$. 
We see a large deviation.
}
\end{figure}%
\begin{figure}[t]
\epsfysize=8cm \epsfbox{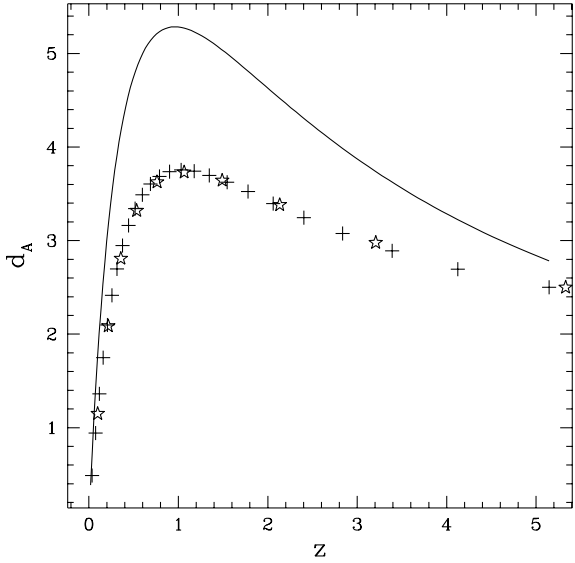}
\caption{Same with Fig.1 for closed models($\Omega = 1.1$).
The solid line 
shows the $\dA$--$z$ relation in a closed
FL model with $\Omega = 1.1, H_\protect{\rm I\protect} = \Hsi$,
$z_\protect{\rm I\protect} = z(i=25)$. 
We see that this FL model does not approximate the dust-shell model.
}
\end{figure}%
\begin{figure}[t]
\epsfysize=8cm \epsfbox{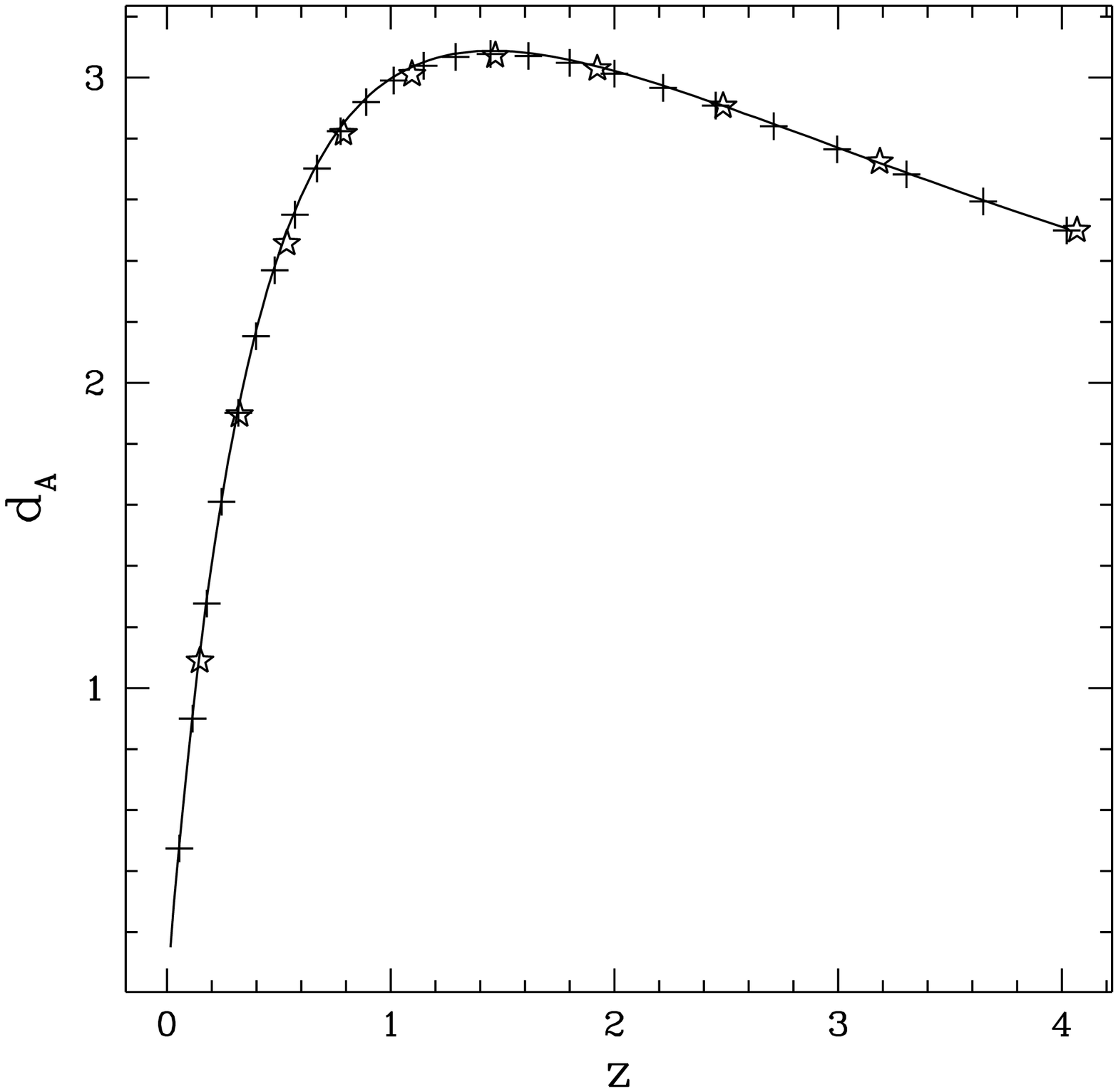}
\caption{Angular diameter distance-redshift relation in the
  open ($\Omega = 0.9$) dust-shell universe  
for choice B. Compare with Fig.3.
The FL relation (solid line)
agrees well with the relation of the  dust-shell model.
}
\end{figure}%
\begin{figure}[ht]
\epsfysize=8cm \epsfbox{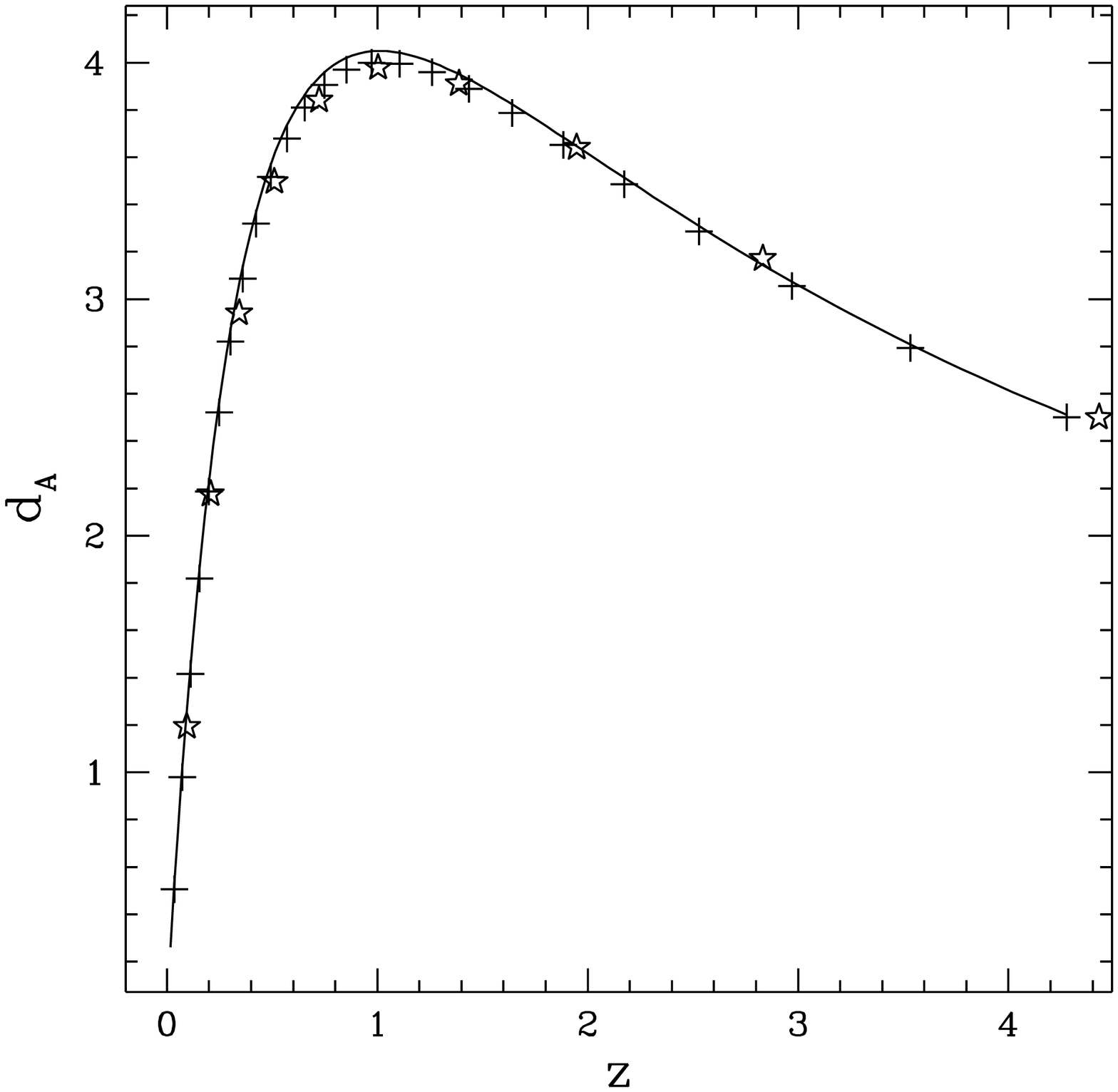}
\caption{Angular diameter distance-redshift relation in the
  closed ($\Omega = 1.1$) dust-shell universe  
for choice B. Compare with Fig.4.
The FL relation (solid line)
agrees well with the relation of the  dust-shell model.
}
\end{figure}
\begin{figure}[b]
\epsfysize=8cm \epsfbox{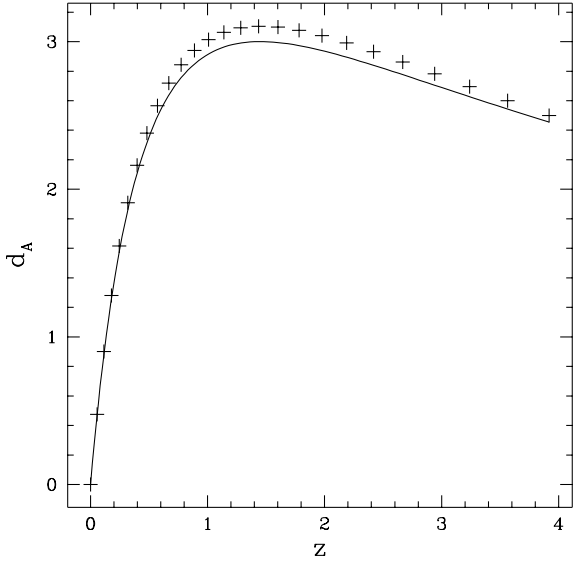}
\caption{Angular diameter distance-redshift relation in the dust-shell universe
for choice C (open model with $\Omega = 0.9$).
We see a mild deviation.}
\end{figure}%

\end{document}